# Gene selection for cancer classification using a hybrid of univariate and multivariate feature selection methods

Min Xu[*]

Lab of Functional Genomics

Institute of Molecular and Cell Biology

30 Medical Drive, Singapore 117609

xumin@ieee.org

Rudy Setiono

School of Computing

National University of Singapore

Lower Kent Ridge Road, Singapore 119260

rudys@comp.nus.edu.sg

___________________________________

[*]To whom correspondence should be addressed.

# ABSTRACT

Various approaches to gene selection for cancer classification based on microarray data can be found in the literature and they may be grouped into two categories: univariate methods and multivariate methods. Univariate methods look at each gene in the data in isolation from others. They measure the contribution of a particular gene to the classification without considering the presence of the other genes. In contrast, multivariate methods measure the relative contribution of a gene to the classification by taking the other genes in the data into consideration. Multivariate methods select fewer genes in general. However, the selection process of multivariate methods may be sensitive to the presence of irrelevant genes, noises in the expression and outliers in the training data. At the same time, the computational cost of multivariate methods is high. To overcome the disadvantages of the two types of approaches, we propose a hybrid method to obtain gene sets that are small and highly discriminative.

We devise our hybrid method from the univariate Maximum Likelihood method (LIK) and the multivariate Recursive Feature Elimination method (RFE). We analyze the properties of these methods and systematically test the effectiveness of our proposed method on two cancer microarray datasets. Our experiments on a leukemia dataset and a small, round blue cell tumors dataset demonstrate the effectiveness of our hybrid method. It is able to discover sets consisting of fewer genes than those reported in the literature and at the same time achieve the same or better prediction accuracy.

# 1   INTRODUCTION

Tissue classification is crucial to cancer diagnosis. It used to be based on morphological appearances, which are often hard to measure and differentiate, so the classification result could be very subjective. The emergence of microarray technology has greatly improved the classification. It is now possible to bring state-of-the-art machine learning methods into the classification process on the data collected at the molecular level. Microarray is a technique for monitoring the expression level of a large number of genes in parallel. Here gene expression refers to the process of transcribing the DNA sequence of a gene into RNA, and the gene expression level indicates how active a gene is in a certain tissue or under a certain experimental condition.

Microarray data can be used in the discovery and prediction of cancer classes. The discovery of previously undefined classes is usually achieved with the use of clustering techniques (Golub et al., 1999; Alizadeh et al., 2000; Bittner et al., 2000; Perou et al., 2000). Examples of these techniques include hierarchical clustering (Weinstein et al., 1997), K-means clustering (Tavazoie et al., 1999) and Self Organizing Maps (Tamayo et al., 1999). In class prediction – where labels are assigned to tissues according to their expression patterns – statistical methods or supervised machine learning methods are useful tools that generally yield high accuracy in predictions. Examples of such methods that have been used in molecular level cancer classification in recent years include the Decision Tree (Cai et al., 2000), Neural Networks (Khan et al., 2001), Support Vector Machine (Furey et al., 2000; Guyon et al., 2002, Mukherjee et al., 2000), and the Naïve Bayesian classifier (Keller et al., 2000).

Prior to classification, it is important to identify relevant genes. In studying the gene expression of a cancer tissue, normally less than a hundred microarray data samples are collected. Each sample consists of expression measurements from thousands to tens of thousands of genes. The measurements are usually noisy: some noises come from the experimental environment, while other noises are from non-uniform genetic backgrounds of the samples being compared (Keller et al., 2000). Although only a small portion of the genes is expected to be relevant to the classification, the combinatorial effect of irrelevant genes has the potential to suppress the contribution of those relevant to the classification (Keller et al., 2000). Thus, the importance of selecting relevant genes before the classification cannot be over-emphasized. The selected genes could provide vital clues in understanding the disease mechanism. Gene selection is important also for the practical reason of reducing the cost of clinical diagnoses. It is much cheaper to focus only on the expression of a few genes rather than on thousands of genes for the diagnoses. At the same time, the resulting classification model could also be expected to be more accurate. It is therefore not surprising that much effort have been put into developing methods for gene selection.

Most gene selection methods fall into two categories: univariate methods and multivariate methods (Liu and Motoda, 1998). Univariate methods (Golub et al., 1999; Slonim et al., 2000; Keller et al., 2000) consider the contribution of individual genes to the classification independently. It does not measure the contribution of a particular gene in relation to the presence of other genes in the data. Multivariate methods such as the Recursive Feature Elimination (RFE) by Guyon et al., (2002) and Optimize Leave-one-out (LOO) by Mukherjee et al., (2000) measure the relative contribution of a gene to the classification by taking other genes into consideration at the same time. Both these methods are wrapper methods (Liu and Motoda, 1998) based on the Support Vector Machine (SVM). They select the set of genes by continuously eliminating genes that have relatively small contribution to classification as measured by the accuracy of the SVM models on the whole gene set.

This paper presents a hybrid method for gene selection and its application to cancer classification problem. Our proposed gene selection method combines the Maximum Likelihood method of Keller et al. (2000) and the Recursive Feature Elimination method based on the Support Vector Machine approach of Guyon et al. (2002). The two gene selection methods are described in detail in Section 2. In Section 3, the advantages and disadvantages of the two methods are compared. We describe how these methods can be combined to further improve classification accuracy. The performance of our hybrid approach is evaluated in Section 4 by comparing the performance of various classification methods trained using the selected genes. We conclude the paper in Section 5.

## 2  PROBLEM DESCRIPTION AND PRIOR WORKS ON GENE SELECTION

### 2.1  *Cancer classification via gene expression*

The quantified microarray data usually consists of the expressions of thousands of genes, comprising a relatively small number of samples of usually less than a hundred. Throughout this paper, we assume that we have a training dataset of *n* labeled samples, each of which is formed by expression of *m* genes, and each gene is a feature for the classification problem. We define the labeled data as follows:

$$\{(\mathbf{x}^1, y^1), \ldots, (\mathbf{x}^n, y^n)\}, \quad \mathbf{x}^i \in \Re^m, \quad \mathbf{x}^i = [x_1^i, \ldots, x_m^i]^T, \quad y^i = \begin{cases} +1, & \text{if sample } i \in \text{ class } a \\ -1, & \text{if sample } i \in \text{ class } b \end{cases}, \quad i = 1, \ldots, n. \quad (1)$$

Here class *a* can be cancerous, and class *b* can be non-cancerous, or they can be two different types of cancer. Cancer classification via gene expression is a general approach applicable to most kinds of cancers. It was first proposed by (Golub et al., 1999). The authors of the paper also proposed a method to rank the usefulness of genes in discriminating leukemia classes based on the mean and standard deviation of their expression. Suppose the expression of the *g* th gene in the training samples for classes *a* and *b* have mean $\mathbf{m}_a^g$ and $\mathbf{m}_b^g$, and standard deviation $\mathbf{d}_a^g$ and $\mathbf{d}_b^g$, respectively. The relevance of the gene for distinguishing between class *a* samples and class *b* samples is ranked according to the correlation

$$\text{corr}^g = \frac{\mathbf{m}_a^g - \mathbf{m}_b^g}{\mathbf{d}_a^g + \mathbf{d}_b^g}. \quad (2)$$

The genes with the most positive or most negative correlations are expected to have the best individual discriminating ability.

Given a selected gene set *G*, Golub et al. (1999) and Slonim et al. (2000) proposed the G-S classification model

$$\text{class}(\mathbf{x}) = \sum_{g \in G} \left( x^g - \frac{\mathbf{m}_a^g + \mathbf{m}_b^g}{2} \right) \times \text{corr}^g. \quad (3)$$

The computed value of class(**x**) is a real number. It takes all genes in *G* into consideration. If it is positive, then the test sample **x** is predicted to belong to class *a*; if it is negative, **x** is predicted to belong class *b*.

## 2.2 Likelihood feature selection method and Bayesian classification

Keller et al. (2000) proposed the Maximum Likelihood gene selection (LIK) method. Denote the event that a sample belongs to class $a$ or class $b$ by $M_a$ and $M_b$, respectively. The difference in the log likelihood is used to rank the usefulness of gene $g$ for distinguishing the samples of one class from the other. The LIK score is computed as follows:

$$LIK^g_{a \to b} = \log(P(M_a \mid x^g_{a,1}, \ldots, x^g_{a,n_a})) - \log(P(M_b \mid x^g_{a,1}, \ldots, x^g_{a,n_a})) \qquad (4)$$

and

$$LIK^g_{b \to a} = \log(P(M_b \mid x^g_{b,1}, \ldots, x^g_{b,n_b})) - \log(P(M_a \mid x^g_{b,1}, \ldots, x^g_{b,n_b})), \qquad (5)$$

where $P(M_i \mid x^g_{j,1}, \ldots, x^g_{j,n_j})$ is *a posteriori* probability that $M_i$ is true given the expression values of the $g$th gene of all the training samples that belong to class $j$, where $n_j$ is the number of training samples that belong to class $j$. According to Bayes rule

$$P(M \mid X)P(X) = P(X \mid M)P(M) \qquad (6)$$

Three assumptions are made by the method. First is the assumption of equal prior probabilities of the classes

$$P(M_a) = P(M_b), \qquad (7)$$

and second is the assumption that the conditional probability of $X$ falls within a small non-zero interval centered at $x$ given $M$ can be modelled by a normal distribution

$$P(x \mid M) = \frac{1}{d\sqrt{2p}} e^{\frac{-(x-m)^2}{2d^2}}, \qquad (8)$$

where $\mathbf{m}$ and $\mathbf{d}$ are the mean and standard deviation of $X$ respectively. The values $\mathbf{m}$ and $\mathbf{d}$ can be estimated from the training data. With the third assumption that the distributions of the expression values of the genes are independent, we obtain the LIK ranking of class $a$ over class $b$ for the $g$ th gene as follows:

$$\begin{aligned} LIK_{a\to b}^g &= \log(P(x_{a,1}^g,\ldots,x_{a,n_a}^g \mid M_a)) - \log(P(x_{a,1}^g,\ldots,x_{a,n_a}^g \mid M_b)) \\ &= \log(\prod_{i=1}^{n_a} P(x_{a,i}^g \mid M_a)) - \log(\prod_{i=1}^{n_a} P(x_{a,i}^g \mid M_b)) \\ &= \sum_{i=1}^{n_a}\left(-\log(\mathbf{d}_a^g) - \frac{(x_{a,i}^g - \mathbf{m}_a^g)^2}{2(\mathbf{d}_a^g)^2} + \log(\mathbf{d}_b^g) + \frac{(x_{a,i}^g - \mathbf{m}_b^g)^2}{2(\mathbf{d}_b^g)^2}\right), \end{aligned} \quad (9)$$

and similarly, the LIK ranking of class $b$ over class $a$ for this gene is

$$\begin{aligned} LIK_{b\to a}^g &= \log(P(x_{b,1}^g,\ldots,x_{b,n_b}^g \mid M_b)) - \log(P(x_{b,1}^g,\ldots,x_{b,n_b}^g \mid M_a)) \\ &= \sum_{i=1}^{n_b}\left(-\log(\mathbf{d}_b^g) - \frac{(x_{b,i}^g - \mathbf{m}_b^g)^2}{2(\mathbf{d}_b^g)^2} + \log(\mathbf{d}_a^g) + \frac{(x_{b,i}^g - \mathbf{m}_a^g)^2}{2(\mathbf{d}_a^g)^2}\right). \end{aligned} \quad (10)$$

Genes that have higher likelihood scores are expected to have better ability to distinguish the classes. Once these genes are selected, the Naïve Bayesian classification method is applied to classify the samples. Given an expression vector $\mathbf{x}$ of $m$ selected genes, the classification of a sample is computed as follows:

$$\text{class}(\mathbf{x}) = \underset{i}{\arg\max}(\log P(M_i \mid \mathbf{x})), \quad i = a, b, \quad (11)$$

where $P(M_i \mid \mathbf{x})$ is *a posteriori* probability that $M_i$ is true given $\mathbf{x}$. Applying the Bayes rule once again, the class for vector $\mathbf{x}$ can be predicted as

$$\text{class}(\mathbf{x}) = \arg\max_i (\log P(\mathbf{x} \mid M_i))$$

$$= \arg\max_i (\sum_{g=1}^{m} \log P(x^g \mid M_i)) \quad (12)$$

$$= \arg\max_i (\sum_{g=1}^{m} -\log \boldsymbol{d}_i^g - \frac{(x^g - \boldsymbol{m}_i^g)^2}{2(\boldsymbol{d}_i^g)^2}), \quad i = a, b.$$

In binary classification, we can be more confident about the classification when the difference between $\log P(\mathbf{x} \mid M_a)$ and $\log P(\mathbf{x} \mid M_b)$ is higher. In order to obtain more information regarding the confidence of the classification, we compute the following

$$\text{class}(\mathbf{x}) = \log P(\mathbf{x} \mid M_a) - \log P(\mathbf{x} \mid M_b)$$

$$= \sum_{g=1}^{m} \left[ -\log \boldsymbol{d}_a^g - \frac{(x^g - \boldsymbol{m}_a^g)^2}{2(\boldsymbol{d}_a^g)^2} \right] - \sum_{g=1}^{m} \left[ -\log \boldsymbol{d}_b^g - \frac{(x^g - \boldsymbol{m}_b^g)^2}{2(\boldsymbol{d}_b^g)^2} \right]. \quad (13)$$

A positive difference leads to the sample being predicted as class $a$, and a negative difference class $b$. The larger the difference, the more confident we are about the prediction. We also make use of this difference when computing another measure of accuracy, i.e. acceptance rate, which will be discussed in Section 4.2.

## 2.3 Recursive Feature Elimination (RFE) by SVM

The Support Vector Machine is rooted in statistical learning theory (Vapnik, 2000). It is different from the Naïve Bayesian classification method in the sense that SVM tries to maximize the separation between samples of two classes. Normally, only a subset of the data samples determines the decision hyperplane. Suppose the $n$ data samples belong to two classes $\{(\mathbf{x}_1, y_1), \ldots, (\mathbf{x}_n, y_n)\}$, $\mathbf{x}_i \in \mathfrak{R}^m$ and $y_i = 1$ or $-1$. A support vector machine tries to find a hyperplane $\mathbf{w}^T \mathbf{x} + b = 0$, which satisfies

$$y_i(\mathbf{w}^T \mathbf{x}_i + b) \geq 1 - \boldsymbol{x}_i, \quad i = 1, \ldots, n, \quad (14)$$

where $x_i \geq 0, i = 1,\ldots,n$ are slack variables. As the distance from a sample to the hyperplane is inversely proportional to $\mathbf{w}^T\mathbf{w}$, a quadratic minimization problem is formulated as follows:

$$\text{minimize} \quad \mathbf{w}^T\mathbf{w} + C\sum_{i=1}^{n} x_i$$

$$\text{subject to} \quad y_i(\mathbf{w}^T\mathbf{x}_i + b) \geq 1 - x_i, \quad i = 1,\ldots,n, \tag{15}$$

where $C$ is a parameter to balance the generalization ability represented in the first term $\mathbf{w}^T\mathbf{w}$, and separation ability indicated in the second term $\sum_{i=1}^{n} x_i$. A smaller value of the first term corresponds to better generalization, while the fewer positive values of the slack variables in the second term correspond to fewer misclassifications on the training samples. When the later is equal to zero, the training samples are linearly separable and there is no misclassification.

Using the Karush-Kuhn-Tucker condition (Bazaraa et al., 1993), the problem is reformulated as the dual quadratic program

$$\text{maximize} \quad W(\mathbf{a}) = \sum_{i=1}^{n} a_i - \frac{1}{2}\sum_{i=1}^{n}\sum_{j=1}^{n} y_i y_j a_i a_j \mathbf{x}_i^T \mathbf{x}_j$$

$$\text{subject to} \quad \sum_{i=1}^{n} y_i a_i = 0, \quad 0 \leq a_i \leq C, \quad i = 1,\ldots,n, \tag{16}$$

where $a_1,\ldots,a_n$ are the Lagrange multipliers associated with the constraints of the quadratic problem (15). The dual problem (16) is a quadratic optimization problem with a unique solution. Those sample vectors with positive Lagrange multipliers are called support vectors. The weight and bias of the optimal separating hyperplane can be easily computed from the solution of the dual problem

$$\mathbf{w}^* = \sum_{i=1}^{n} a_i^* y_i \mathbf{x}_i \tag{17}$$

and

$$b^* = 1 - \mathbf{w}^{*T}\mathbf{x}_i, \quad \text{if } y_i = +1 \tag{18}$$

or

$$b^* = \mathbf{w}^{*T}\mathbf{x}_i - 1, \quad \text{if } y_i = -1. \tag{19}$$

The absolute values of the weights of a trained SVM can be used to indicate the relevance of the corresponding features to classification and the ranking of each of the weights takes all the other features into consideration.

Guyon et al. (2002) proposed a Recursive Feature Elimination (RFE) method, which is a greedy wrapper feature selection method (Kohavi and John, 1997). It iteratively trains new SVMs and eliminates the features whose corresponding absolute weight is the smallest from the dataset. This approach to feature elimination is similar to the backward feature selection approach in statistical multiple linear regression. Different strategies for eliminating the features are recommended; they can be eliminated one by one, 10 percent at a time, or half at a time. The solution set would be different with different strategies.

The MATLAB-like pseudo-code of the algorithm that removes one feature at a time is as follows:

% The function RFE accepts a vector of feature vectors $\mathbf{S}$, and class label vector $\mathbf{y}$.

$\text{RFE}(\mathbf{S} = [\mathbf{f}_1,\ldots,\mathbf{f}_m], \mathbf{y})$

While $\text{length}(\mathbf{S}) \geq 1$

% Train SVM with the remaining features in $\mathbf{S}$, and obtain the weights of trained SVM.

$\mathbf{w} = \text{svm\_training}(\mathbf{S}, \mathbf{y})$

% Find the $f$ th weight that has the minimum magnitude.

$f = \arg\min(\text{abs}(w_i)), i = 1,\ldots,\text{length}(\mathbf{S})$

% Remove the $f$ th element in $\mathbf{S}$.

$\mathbf{S} = [\mathbf{S}(1:f-1) \quad \mathbf{S}(f+1:\text{length}(\mathbf{S}))]$

end

## 3  IMPROVING SELECTION BY COMBINING LIK AND RFE

LIK is a univariate ranking method while RFE is a multivariate ranking method based on SVM. The separating hyperplane obtained by SVM divides samples from two different classes and its coefficient weights reflect the relative contribution of the corresponding genes to the classification as a whole. In cancer classification problems, however, the sample size is usually very small compared with the number of genes. Moreover, many genes are expected to contribute to classification together with some other genes rather than contributing independently, which makes multivariate ranking more likely to choose a smaller number of genes with higher prediction accuracy than a univariate ranking. However, most of the genes in a microarray dataset are expected to be irrelevant to the classification. The expression of many irrelevant genes may obscure the discriminative information of the relevant genes. This can be seen in the formulation of the SVM dual problem where the coefficients of the quadratic terms in the dual problem are computed as the scalar products of two inputs

$$\mathbf{x}_i^T \mathbf{x}_j = \sum_{k=1}^{m} x_{ik} x_{jk} \tag{20}$$

The gene elimination process is very sensitive to change in the feature set. SVM also has the disadvantage that it is sensitive to outliers as discussed in Guyon et al. ( 2002). In microarray data, the outliers may be introduced by: 1) noise in the expression data, or 2) incorrectly identified/labeled samples in the training dataset. It is therefore more beneficial to apply RFE on a dataset with a reduced number of features. A univariate feature selection algorithm can be used to first efficiently reduce the large number of features originally present in the dataset and a multivariate feature selection method such as RFE can then be applied to remove more features. To summarize, we first identify and remove genes that are expected to have low discrimination ability as indicated by LIK scores. Then, we apply RFE to reduce the size of the feature set further. With this integrated approach to feature selection, we are able to achieve good classification performance with fewer genes than those reported by Guyon et al. (2002) and Keller et al. (2000).

A multivariate method is usually very time-consuming when applied to a dataset with thousands of genes. For RFE, in order to eliminate one or more genes, a new SVM has to be trained, and the overall computational cost is $\Omega(m^2 n^2)$. On the other hand, LIK ranks genes independently, which makes the computational complexity of LIK, $O(mn)$. Using LIK first to reject a large number of genes, and then using RFE to perform further selection will save significant running time compared to just using RFE alone. This is especially important when the improvement of microarray technology makes it possible to obtain gene expression values from tens or even hundreds of thousands of genes.

## 4    EXPERIMENTAL RESULT

### *4.1    Dataset description*

We tested our proposed method for feature selection on two datasets. The first dataset was from a human acute leukemia microarray sample (Golub et al., 1999). The dataset contained 72 samples; each sample consisted of 7129 genes and belonged to either acute myeloid leukemia (AML) or acute lymphoblastic leukemia (ALL). There were 38 training samples (27 ALL and 11 AML samples) and 34 test samples (20 ALL and 14 AML samples).

The second dataset was collected from small, round blue cell tumors (SRBCT) (Khan et al., 2001). There were 88 samples altogether, each of which was described by 2308 genes. The SRBCT samples were divided into four classes: neuroblastoma (NB), rhabdomyosarcoma (RMS), Burkitt lymphomas (BL) and the Ewing family of tumors (EWS). Because we used methods for binary classification problems, we decomposed the problem of distinguishing samples from four classes into four separate binary classification problems. Khan et al. divided the dataset into 63 training samples (23 EWS, 8 BL, 12 NB, 20 RMS) and 25 test samples (6 EWS, 3 BL, 6 NB, 5 RMS, 5 non-SRBCT including 2 Sk.Muscle, 2 Sarcoma and 1 Prostate).

The goal of our experiment was to try to select the smallest set of genes that could achieve a sufficiently high classification performance using a combination of LIK and RFE. The classification algorithms used were the Support Vector Machine and the Naïve Bayesian method.

### 4.2 Performance measures

We ran our experiments on a Pentium 4 1.4GHz computer with 512-megabyte memory. We wrote and ran our program using MATLAB 6.1. The support vector machine was constructed with the Support Vector Machine Toolbox from http://theoval.sys.uea.ac.uk/~gcc/svm/toolbox which was developed by Gavin Cawley. For the SVM, we set $C = 100.0$ and used the linear kernel, the same as those used by Guyon et al. (2002).

We employed two performance measures. Suppose there were $n$ samples predicted with output $o_1, \ldots, o_n$, and their corresponding class labels were $y_1, \ldots, y_n$. If the prediction output of a classifier for a sample had the same sign as that of its true class, we considered the sample to be correctly classified. The first performance measure was accuracy, i.e. the number of correctly classified samples over the total number of test samples, that was

$$\text{accuracy} = \frac{\left|\{i \mid o_i y_i > 0, i = 1, \ldots, n\}\right|}{n}, \quad (21)$$

where $|S|$ denoted the cardinality of the set $S$.

The second performance measure was the acceptance rate, which assessed the performance more strictly. The acceptance rate was computed as follows:

$$\text{acceptance rate} = \frac{\left|\left\{i \;\middle|\; o_i y_i > -\min_{j=1,\ldots,n}(o_j y_j), i = 1, \ldots, n\right\}\right|}{n}. \quad (22)$$

In the tables and figures in the next sections, we will denote accuracy and acceptance rate by acu and acp, respectively. Obviously, the acceptance rate cannot be higher than accuracy. For a more detailed illustration of the two measures, see (Guyon et al., 2002). Because the number of samples was very small, we used the leave-one-out method for validating the classifier on training samples as well as on all samples. When there were $n$ samples, leave-one-out was a technique to iteratively choose each sample for testing, and the remaining samples for training. A total of $n$ classifiers were trained, and $n$ predictions were made. The accuracy and acceptance rate were computed from the predictions and labels of the corresponding test samples.

### *4.3 Results*

LIK has been shown to outperform the G-S method (Keller et al, 2000), while RFE has been shown to be better than the G-S method (Guyon, et al., 2002). Here we show that the combination of LIK and RFE can achieve better results than using LIK or RFE alone.

Before applying the hybrid LIK+RFE, we calculated the mean and standard deviation of the expression values of each gene in the training dataset. We then normalized the gene expression values of all samples in both the training and test datasets by subtracting the mean value and then dividing the difference by the corresponding standard deviation. The same normalization procedure was also followed by Golub et al. (1999).

**4.3.1  Leukemia dataset**

Figure 1 shows the sorted LIK scores. We chose equal numbers of genes with the highest $LIK_{ALL \to AML}$ and $LIK_{AML \to ALL}$ score as the initial gene sets for RFE. We plotted in this figure the scores of the top (2 x 80) genes. The top $i$ th gene according to $LIK_{ALL \to AML}$ always has a higher $LIK_{ALL \to AML}$ value than the corresponding top $i$ th gene's $LIK_{AML \to ALL}$ score. Genes with low LIK scores are not expected to be good discriminators. We decided to pick the top genes to check their discriminating ability. In particular, we ran experiments using the top (2 x 10), (2 x 20), (2 x 30) genes. We found the best performance was obtained when 2 x 20 top ranking genes were selected. The performance was measured by computing the prediction accuracy and acceptance rate of SVM and Bayesian classifiers built using the selected genes on the test samples.

Figure 2 shows the accuracy and acceptance rate using two different experimental settings: leave-one-out and train-test split. For the leave-one-out (LOO) setting, we computed the performance measures using only the 38 training samples as well as on the entire dataset consisting of 72 samples. For the train-test split, the measures shown were computed on the 34 test samples, while the measures on the 38 training samples are not reported in the table. A series of experiments were conducted to find the smallest number of genes that would give good performance measure. The experiments started with all 40 (= 2 x 20) genes selected by the LIK feature selection. One gene at a time was eliminated using RFE. RFE feature selection was conducted until there was only one gene left. For a selected subset of genes, the performance measures were computed under all experimental settings and using both SVM and Bayesian classifiers.

As can be seen from the figure, the SVM classifier achieved almost perfect accuracy and acceptance rate when there were three to 14 genes used to find the separating hyperplane. On the other hand, when the Naïve Bayesian method was used for classification, almost perfect performance was achieved with as many as 40 genes in the model. Elimination of the genes by RFE one by one showed that the results could be maintained as long as there are at least three genes in the model. This stability in performance indicates the robustness of the RFE feature selection method when given a pre-selected small subset of relevant genes, as identified by the LIK method. It is worth noting that the acceptance rate on the test samples was almost constant with at least three genes, both when the SVM classifier and the Naïve Bayesian classifier were used for prediction. We emphasize here that the hybrid LIK+RFE feature selection was run using the 38 training samples; the classifiers were also built using the same set of training samples without the use of any information from the data in the test set.

A set of three genes was discovered to give perfect accuracy and acceptance rate regardless of the experimental settings and the classifiers used. These genes are listed in Table 1. They have also been identified as relevant genes in this dataset by several researchers. Golub et al. (1999) identified U05259_rna1_at and M27891_at as relevant, while Keller et al. (2000) identified the gene X03934_at as relevant. On the other hand, Guyon et al. (2002) identified a completely different set consisting of four genes.

Since there were only three genes selected by the hybrid LIK+RFE method, we are able to visualize the distribution of both the training and test samples in a three-dimensional space. Figure 3 shows the plot of the samples. In this figure, we differentiate between acute myeloid leukemia (AML) and acute lymphoblastic leukemia (ALL) samples. There were actually two different types of ALL samples. These were B-cells or T-cells as determined by whether they arose from a B or a T cell lineage (Keller et al., 2000). From the figure, we can see that all except one B-cell sample had almost constant expression values for two genes, namely M27891_at and X02934_at. Training sample number 17 was the one ALL B-cell that was an outlier. On the other hand, all T-cell samples had almost constant expression values for genes U05259_mal_at and M27891_at, while all AML samples had similar expression values for U05259_mal_at and X03934_at. The plot shows that the three selected genes were also useful in differentiating ALL B-cell and T-cell samples.

For comparison purposes, the classification performance of SVM and Naïve Bayesian classifiers built using genes selected according to their LIK scores only is shown in Figure 4. For the results shown in this figure, we started with the same set of 40 genes and removed one gene at a time according to their LIK scores. As can be seen from the figure, the results were not as good as those shown in Figure 2. In particular, using SVM classifiers, the accuracy and the acceptance rate were more than 80 percent when there were still more than 20 genes in the model. The acceptance rate drops drastically when there are fewer genes. Naïve Bayesian classifiers performed well when there were more than 21 genes. Further removal of more genes according to their LIK scores caused the acceptance rate to drop considerably. When there were fewer than five genes, the accuracy and the acceptance rate of the classifiers were low.

The performance of SVM and Naïve Bayesian classifiers built using the genes selected by RFE is depicted in Figure 5. We started with all 7129 genes in the feature set. We built an SVM using the training samples with expression values of all the genes and measured its performance on the test samples. We also built a Naïve Bayesian classifier and measured its performance as well. The gene that had the smallest absolute weight in the SVM-constructed hyperplane was removed, and the process of training and testing was repeated with one fewer gene. This process was continued until there were no more genes to be removed.

An interesting point to note from the results depicted in Figure 5 is the sharp improvement in the acceptance rate of the Bayesian classifiers when the number of genes was reduced from 2773 to 2772. The gene that was eliminated at this stage was M26602_at. The acceptance rates stayed at 100 percent when there were 2772 to 1437 genes. Further removal of genes caused the rate to deteriorate gradually. On the other hand, the performance of SVM was more stable. With more than 519 genes, both the accuracy and the acceptance rate were at least 90 percent.

We also experimented with choosing the top genes according to their LIK scores. We selected genes with LIK scores that were higher than a certain threshold. The threshold values tested were 1500, 2000, and 2500. Note that there were always more genes selected because of their high $LIK_{ALL \to AML}$ than genes selected because of their $LIK_{AML \to ALL}$ values. The best performance was obtained when the threshold was set to 1500. A total of 62 genes met this threshold value and were used to form the initial gene set for RFE. After applying RFE, we obtained a set of four genes that achieves perfect accuracy and acceptance rate on the training and test samples under all three experimental settings. The set of four selected genes is shown in Table 2. Two out of the four genes were the same ones as those selected using the (2 x 20) top initial genes listed in Table 1. These genes were U0529_rnal_at and M27891_at. The genes M16336_s_at was also found by Keller at al. (2000) to be an important gene for classification.

The performance of the SVM classifiers with genes selected using just the RFE approach was slightly different from that reported by Guyon et al. (2002). The reason for this could be the variation in the implementation of the quadratic programming solvers. The Matlab toolbox uses Sequential Minimal Optimisation algorithm (Platt, 1999), while Guyon et al. used a variant of the soft-margin algorithm for SVM training (Cortes, 1995). Our hybrid LIK+RFE method achieved better performance than other methods reported in the literature. To achieve perfect performance, the RFE implementation of Guyon et al. needed eight genes. When the number of genes was reduced to four, the leave-one-out results on the training samples using SVM achieved only 97 percent accuracy and 97 percent acceptance rate. SVMs trained on 38 training samples with the four selected genes achieved only 91 percent accuracy and 82 percent acceptance rate on the test samples.

Using the genes selected according to their LIK scores and applying the Bayesian method, Keller et al. (2000) achieved 100 percent prediction with more than 150 genes. Hellem and Jonassen (2002) required 20 to 30 genes to obtain accurate prediction by ranking pair-wise contribution of genes to the classification. The classification of the samples was obtained by applying k-nearest neighbours, diagonal linear discriminant and Fisher's linear discriminant methods. Guyon et al. also mentioned the performance of other works on this dataset (Mukherjee et al., 2000; Chapelle et al., 2000; Weston et al., 2001). None of these works reported performance results that are as good as ours.

### 4.3.2 SRBCT dataset

We obtained the expression ratio data of Khan et al. (2001). Before we conducted our experiments, the expression values were transformed by computing their logarithmic values. Base 2 log transformation was used, as this is the usual practice employed by researchers analyzing micraoarray data. In Figure 6, the plot of the gene ranking according to their LIK scores is shown. The LIK scores were computed for differentiating EWS samples from non-EWS samples. The set of top 20 genes according to their $LIK_{EWS \rightarrow Non-EWS}$ ranking contained eight genes that were also in the set of top 20 genes according to their $LIK_{Non-EWS \rightarrow EWS}$ ranking. Hence, when RFE was applied to further eliminate genes from the feature set, it started with 32 unique genes.

For the other three classification problems, the plots would look very similar to Figure 6 and are not shown in this paper. For each of the four problems, LIK selected the top (2 x 20) genes. The number of unique genes selected by LIK and the results of the experiments from solving four binary classification problems are summarized in Table 3. The numbers of unique genes selected by LIK and the smallest numbers of genes required to achieve near perfect performance during the gene elimination process by RFE are shown in the second column of the table. For the three classification problems to identify EWS, BL and NB, the accuracy and the acceptance rates were at least 98 percent for all experimental settings. Those perfect performance results are highlighted in the table. For the fourth classification problem to differentiate between RMS and non-RMS samples, the accuracy rates were at least 92 percent. However, the acceptance rate on the test samples dropped to eight percent for SVM classifier and 16 percent for Naïve Bayesian classifier, respectively.

The poor acceptance rate obtained when predicting RMS test samples suggests that the differences in the output of the classifiers and the actual target values were high for the incorrectly predicted samples. In order to verify the predictions, we plotted the distribution of the samples according to the expression values of three genes, ImageID784224 (fibroblast growth factor receptor 4), ImageID796258 (sarcoglycan, alpha), and ImageID1409509 (troponin T1). The three were selected because their corresponding SVM weights were the largest. The plot is shown in Figure 7. We can clearly see that the two incorrectly classified non-RMS samples were outliers with large values for ImageID1409509 (troponin T1). These two outliers were Sk. Muscle samples TEST-9 and TEST-13, which were misclassified as RMS samples. It should be noted that there were no Sk. Muscle samples in the training dataset.

The genes selected by the hybrid LIK+RFE for each of the four classification problems are listed in Table 4. For the problem of differentiating EWS from non-EWS samples, our method selected five genes, all of which were also selected by Khan et al. (2001). On the other hand, to differentiate between NB and non-NB samples, only three genes were needed and none was selected by Khan et al. All together, the hybrid LIK+RFE identified 15 important genes. This number compares favorably with the total of 96 genes selected by the PCA (Principle Component Analysis) approach of Khan et al.

We also tested the classification performance of SVM and Naïve Bayesian classifiers on genes selected based purely on their LIK scores. For comparison purpose, for each of the four problems, the number of genes was set to be the same as the corresponding final number selected by the hybrid LIK+RFE shown in Table 3. Table 5 summarizes the results. For three of the four classification problems, the performance of the classifiers was not as good as the results reported in Table 3. The accuracy and acceptance rates dropped to as low as 52 percent. The most unexpected results came from the fourth problem to differentiate between RMS and non-RMS samples. The SVM classifier achieved perfect accuracy and acceptance rate using four genes, while the Naïve Bayesian classifier managed to obtain at least 92 percent accuracy and acceptance rate. The four genes were ImageID461425 (MLY4), ImageID784224 (fibroblast growth factor receptor 4), ImageID296448 (insulin-like growth factor 2), and ImageID207274 (Human DNA for insulin-like growth factor II). All these genes were among the 96 genes identified by Khan et al. (2001). Of these four, only one was selected by LIK+RFE, that is, ImageID784224.

In comparison, Khan et al. (2001) used neural networks for multiple classifications to achieve 93 percent EWS, 96 percent RMS, 100 percent BL and 100 percent NB diagnostic classification performance on the 88 training and test samples. Since there were four classes of training data samples, each neural network had four output units. The target outputs were binary encoded, for example, for an EWS sample the target was (EWS=1, RMS=NB=BL=0). A total of 3750 neural networks calibrated with 96 genes were required. The highest average output value from all neural networks determined the predicted class of a new sample. The Euclidean distance between the average values and the target values was computed for all samples in order to derive the probability distribution of the distances. A test sample would be diagnosed as a member of one of the four classes based on the highest average value given by the neural networks. This was provided that the distance value falls within the $95^{th}$ percentile of the corresponding distance probability of the predicted class. Otherwise, the diagnosis would be rejected and the sample would be classified as a non-SRBCT sample. Of the 88 samples in the training and test datasets, eight were rejected. Five of these were non-SRBCT samples in the test set, while the other three actually belonged to the correct class but their distances lay outside the threshold of the $95^{th}$ percentile.

In order to visualize the distribution of the samples based on the expression values of the selected genes, we performed clustering of the genes using the EPCLUST program (http://ep.ebi.ac.uk/EP/EPCLUST). The default setting of the program was adopted; the average linkage clustering and uncentered correlation distance measure were used. Figure 8 shows the clusters. It can be seen clearly from this figure that there existed four distinct clusters corresponding to the four classes in the data. Most of the samples of a class fell into their own corresponding clusters. The five non-SRBCT samples lay between clusters. We conjecture that samples between clusters might not belong to any classes found in the training dataset. Two between-cluster samples, RMS-T7 and TEST-20 were exceptions. RMS-T7, which was nearer to the two Sk. Muscle samples TEST-9 and TEST-13 was actually an RMS sample. TEST-20, which was nearer to Prostate sample TEST-11 than to EWS cluster was actually an EWS sample. These exceptions were consistent with the neural network prediction results of Khan et al. (2001) as the neural networks predicted TEST-9 and TEST-13 to be RMS class, and they predicted TEST-20 and TEST-11 to be EWS class. Both predictions, however, did not meet the $95^{th}$ percentile distance criterion and were therefore rejected. This indicated that these samples were also difficult to differentiate by the neural networks. Different results from our clustering and the neural network classification can be seen for test sample TEST-3, a non-SRBCT sample. The clustering placed TEST-3 between BL and NB clusters. But the neural networks predicted this sample as an RMS sample without meeting the 95$^{th}$ percentile distance criterion.

## 5    CONCLUSION AND DISCUSSION

Microarray is an important tool for cancer classification at the molecular level. It monitors the expression levels of large number of genes in parallel. With large amount of expression data obtained through microarray experiments, suitable statistical and machine learning methods are needed to search for genes that are relevant to the identification of different types of cancer tissues. In this paper, we have proposed a hybrid gene selection method, which combines a univariate selection method, i.e. statistics-based likelihood method LIK, and a multivariate selection method, i.e. the machine learning method RFE, to achieve high classification performance using fewer selected genes. The computational time of this hybrid method is significantly less than that of the multivariate method RFE.

Two datasets were used to systematically study the performance of the hybrid gene selection method. The test on the benchmark leukemia dataset achieved performance results that are as good as or better than those obtained by other researchers. The test on the SRBCT dataset achieved prediction performance that is similar to that reported by other researchers. For both datasets, however, our hybrid gene selection method identified very few genes that are excellent discriminators for the different classes of data samples. For the leukemia dataset, both SVM and Naïve Bayesian classifiers built using these genes achieve perfect accuracy and acceptance rate with only three genes identified. For the SRBCT dataset, a total of 15 genes were found to be sufficient to differentiate the samples by our hybrid method. On the other hand, a PCA method identified 96 genes and the neural network models trained using these selected genes achieved performance that was similar to ours. We believe that the proposed hybrid method has great potential for effectively selecting small discriminative gene sets from microarray data.

Most of the genes selected by our hybrid LIK + RFE method have some relevance to cancer according to literature search in PubMed, a document retrieval service of the National Library of Medicine of United States. However, biological experiments need to be done for further validation of the role of these genes. The performance of the method is also data dependent, as demonstrated in the significant difference in the acceptance rate of the classifiers for the first three binary classification problems and the fourth problem in the SRBCT dataset. Overall, we observe that the classification performance on the test set generally does not change much with the consecutive elimination of a few genes. The removal of one gene would not normally cause a drastic change in the performance of the classifier. Significant drops in accuracy and/or the acceptance rate is observed most frequently when a gene is removed from the optimal set.

The main idea presented in this paper is to combine univariate and multivariate methods for gene selection. We have experimented with various combinations of other feature selection methods. However, none of those combinations yielded better results. These other feature selection methods that we have tried include methods that are based on information gain (Quinlan, 1993), those that make use of neural networks (Setiono and Liu, 1997), those that compute extremal margin (Goyun et al. 2002), and Fisher's linear discriminator.

The study on linear separability (Cover, 1965) implies that when the number of samples is small compared with the number of features, it is possible to find a number of subsets of features that can perfectly distinguish all samples. Our experiments on the leukemia dataset also support this hypothesis: we found two different gene sets consisting of just three or four genes, which can achieve perfect classification performance. Biological study shows that although many genes do not have direct relevance to the cancer under study, their expression may have subtle and systematic difference in different classes of tissues (Alon et al., 1999). Hence, a new challenge for cancer classification arises: to find as many as possible small subsets of genes that can achieve high classification performance. Using only microarray data with these subsets of genes, we can build different classifiers and look for those that have desirable properties such as extremal margin, i.e. wide difference between the smallest output of the positive class samples and the largest output of the negative class samples. Another property could be median margin, which is the difference between the median output of the positive class samples and the median output of the negative class samples. Exhaustively enumerating and evaluating all the gene combinations is computationally NP-hard (non-deterministic Polynomial-time hard) and is feasible only when the number of relevant genes is relatively very small.

Due to its cost, microarray experiments conducted for identifying the genes that are crucial for cancer diagnosis are still scarce. The measurements obtained from the experiments are noisy. These facts make the selection of different sets of relevant genes vital. Moreover, cancer is a complex disease. It is not caused by only a few genes, but also by many other factors (Kiberstis and Roberts, 2002). So even the best selected subsets may not actually be the most crucial ones to the cancer under study. They can, however, be important candidates for a further focused study on the gene interactions within individual subsets, and the relationship between these interactions and the disease. There has been work done on the second order selection. For example, Goyun et al. (2002) found a gene pair that could have zero leave-one-out error on the training samples, but achieved poor performance on test samples. Hellem and Jonassen (2002) also evaluated the contribution of pairs of genes to the classification for the ranking of genes, but they still have to combine multiple pairs of genes to perform classification. We plan to work on finding better ways to develop methods for high order feature selection that would allow the classifiers to achieve high performance with different small sets of genes.

**ACKNOWLEDGEMENTS**

We thank Isabelle Guyon for providing the supplementry data. We also thank Lirong Cui, Yang Wang, Oilian Kon, Jinrong Peng, Wolfgang Hartmann, and Guozheng Li for their numerous helpful consultations.

**TABLES**

| Gene accession number | Description |
|---|---|
| U05259_rna1_at | MB-1 gene |
| M27891_at | CST3 Cystatin C (amyloid angiopathy and cerebral hemorrhage) |
| X03934_at | GB DEF = T-cell antigen receptor gene T3-delta |

**Table 1.** The smallest gene set found that achieves prefect classification performance.

| Gene assection number | Description |
|---|---|
| U05259_rna1_at | MB-1 gene |
| M16336_s_at | CD2 CD2 antigen (p50), sheep red blood cell receptor |
| M27891_at | CST3 Cystatin C (amyloid angiopathy and cerebral hemorrhage) |
| X58072_at | GATA3 GATA-binding protein 3 |

**Table 2.** The genes selected by the hybrid LIK+RFE method. The genes that have LIK scores of at least 1500 were selected initially. RFE was then applied to select these four genes that achieved perfect performance.

| | | SVM | | | | | | Bayesian | | | | | |
|---|---|---|---|---|---|---|---|---|---|---|---|---|---|
| | | Leave-one-out on training samples | | Prediction on test samples | | Leave-one-out on all samples | | Leave-one-out on training samples | | Prediction on test samples | | Leave-one-out on all samples | |
| Classification problem | Initial/final number of genes | Acu | Acp | Acu | Acp | Acu | Acp | Acu | Acp | Acu | Acp | Acu | Acp |
| EWS vs non-EWS | 32/5 | 1.00 | 1.00 | 1.00 | 1.00 | 0.99 | 0.99 | **1.00** | **1.00** | **1.00** | **1.00** | **1.00** | **1.00** |
| BL vs non-BL | 37/3 | **1.00** | **1.00** | **1.00** | **1.00** | **1.00** | **1.00** | 1.00 | 1.00 | 1.00 | 1.00 | 1.00 | 1.00 |
| NB vs non-NB | 34/3 | **1.00** | **1.00** | **1.00** | **1.00** | **1.00** | **1.00** | 0.98 | 0.98 | 1.00 | 1.00 | 1.00 | 1.00 |
| RMS vs non-RMS | 34/4 | 1.00 | 1.00 | **0.92** | **0.08** | 0.99 | 0.88 | 1.00 | 1.00 | **0.92** | **0.16** | 0.97 | 0.35 |

**Table 3.** Experimental results for the SRBCT dataset using the hybrid LIK+RFE.

| Classification problem | Reported by Khan et al., (2001) | Image ID | Description |
|---|---|---|---|
| EWS vs non-EWS | Y | 377461 | caveolin 1, caveolae protein, 22kD |
|  | Y | 295985 | ESTs |
|  | Y | 80338 | selenium binding protein 1 |
|  | Y | 52076 | olfactomedinrelated ER localized protein |
|  | Y | 814260 | follicular lymphoma variant translocation 1 |
| BL vs non-BL | Y | 204545 | ESTs |
|  |  | 897164 | catenin (cadherin-associated protein), alpha 1 (102kD) |
|  | Y | 241412 | E74-like factor 1 (ets domain transcription factor) |
| NB vs non-NB |  | 45632 | glycogen synthase 1 (muscle) |
|  |  | 768246 | glucose-6-phosphate dehydrogenase |
|  |  | 810057 | cold shock domain protein A |
| RMS vs non_RMS |  | 897177 | phosphoglycerate mutase 1 (brain) |
|  | Y | 784224 | fibroblast growth factor receptor 4 |
|  | Y | 796258 | sarcoglycan, alpha (50kD dystrophin-associated glycoprotein) |
|  | Y | 1409509 | troponin T1, skeletal, slow |

**Table 4.** The genes selected by the hybrid LIK+RFE for the four binary classification problems.

| Classification problem | Number of genes | SVM | | | | | | Bayesian | | | | | |
|---|---|---|---|---|---|---|---|---|---|---|---|---|---|
| | | Leave-one-out on training samples | | Prediction on test samples | | Leave-one-out on all samples | | Leave-one-out on training samples | | Prediction on test samples | | Leave-one-out on all samples | |
| | | Acu | Acp | Acu | Acp | Acu | Acp | Acu | Acp | Acu | Acp | Acu | Acp |
| EWS vs non-EWS | 5 | 1.00 | 1.00 | 0.92 | 0.88 | 0.95 | 0.88 | 0.98 | 0.97 | 0.84 | 0.84 | 0.95 | 0.86 |
| BL vs non-BL | 3 | 0.95 | 0.92 | 0.88 | 0.88 | 0.97 | 0.83 | 0.98 | 0.98 | 0.88 | 0.76 | 0.93 | 0.88 |
| NB vs non-NB | 3 | 0.95 | 0.92 | 0.84 | 0.76 | 0.97 | 0.86 | 0.97 | 0.97 | 0.80 | 0.52 | 0.95 | 0.92 |
| RMS vs non-RMS | 4 | **1.00** | **1.00** | **1.00** | **1.00** | **1.00** | **1.00** | 0.97 | 0.95 | 0.92 | 0.92 | 0.97 | 0.95 |

**Table 5.** The performance of SVM and Naïve Bayesian classifiers built using the top genes selected according to their LIK scores.

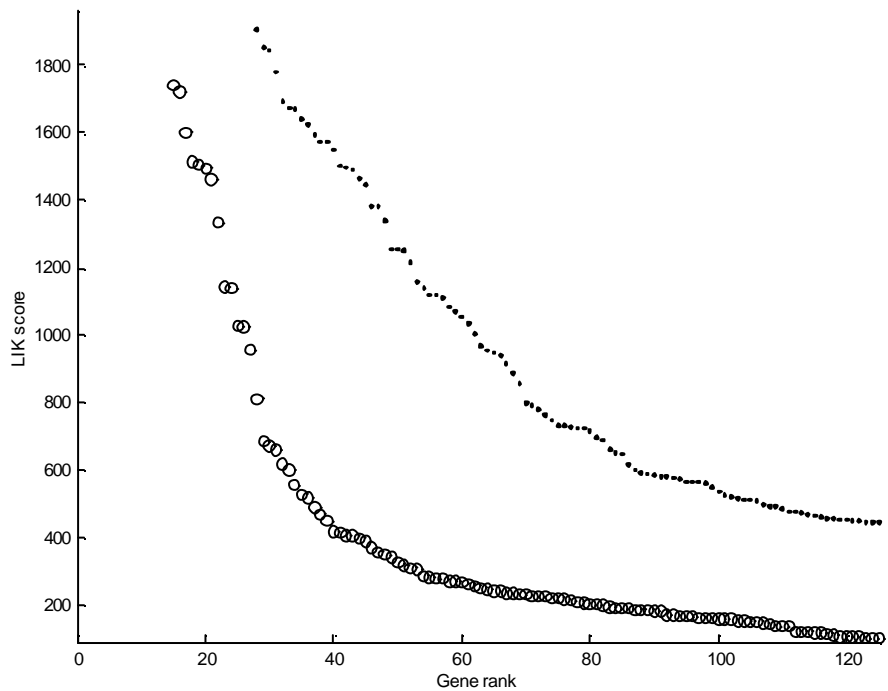

**Figure 1.** Sorted LIK score of a subset of genes in the leukemia dataset. Dots indicate $LIK_{ALL \to AML}$ scores and circles indicate $LIK_{AML \to ALL}$. The top 28 genes according to their $LIK_{ALL \to AML}$ values have scores between 92014 and 1978.3; and the top 15 genes according to their $LIK_{ALL \to AML}$ values have scores between 22852 and 2148.7; they are not shown in this figure.

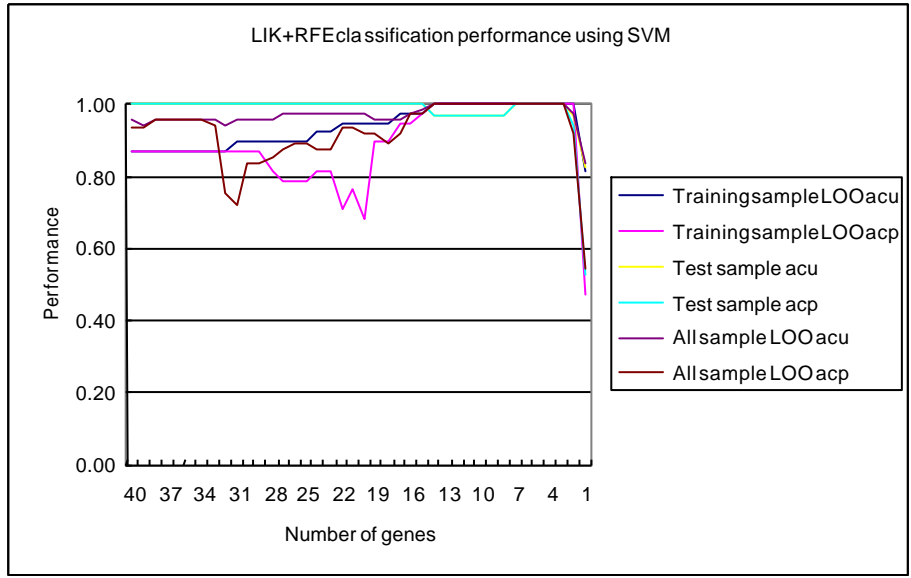

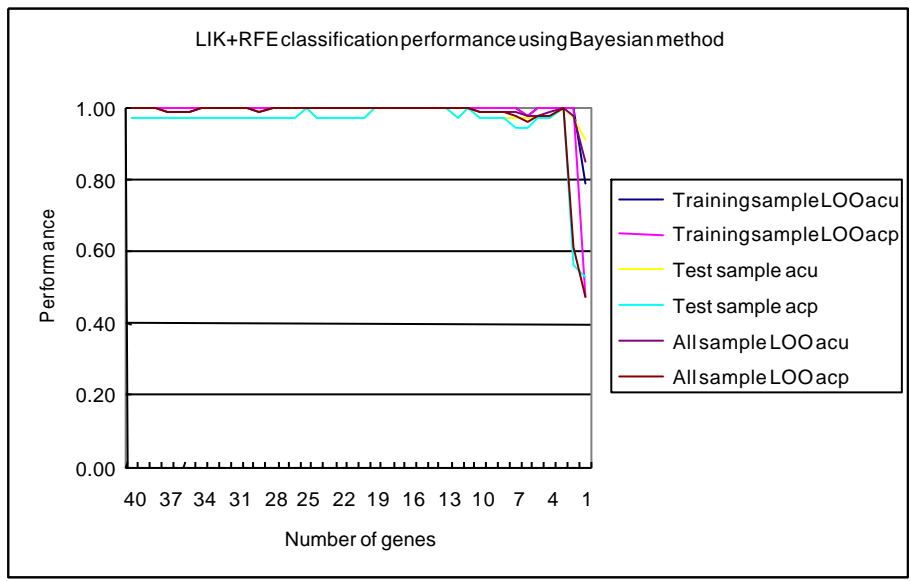

**Figure 2.** Classification performance of genes selected using the hybrid LIK+RFE.

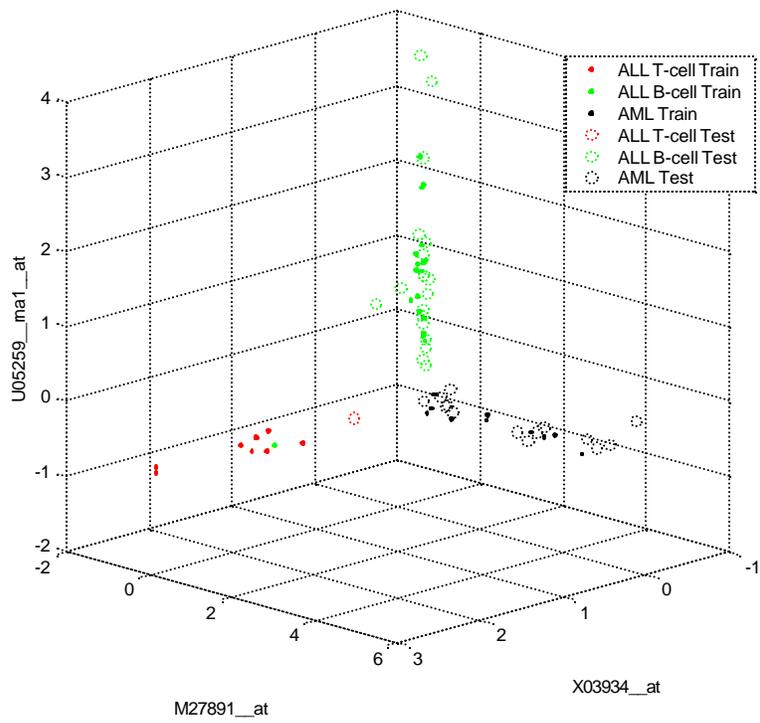

**Figure 3.** Plot of the leukemia data samples according to the expression values of the three genes selected by the hybrid LIK+RFE.

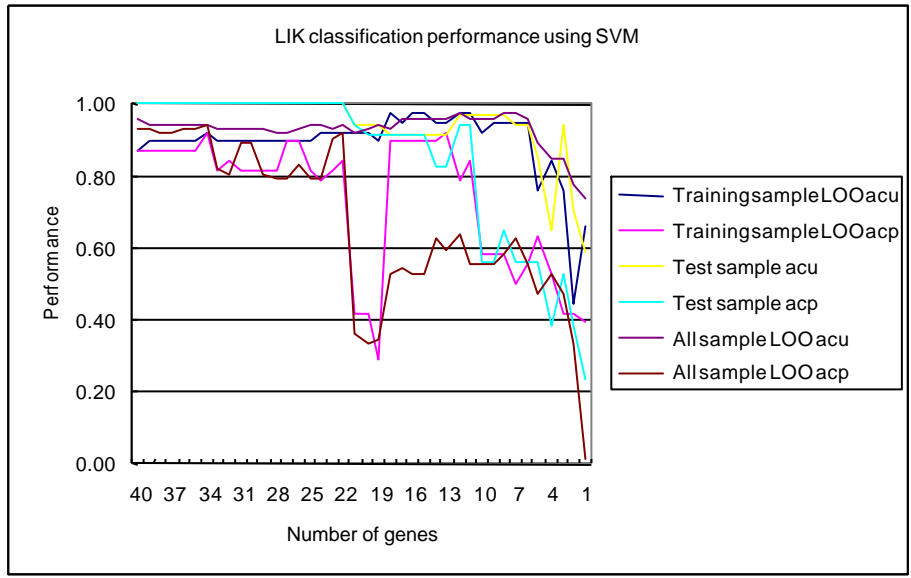

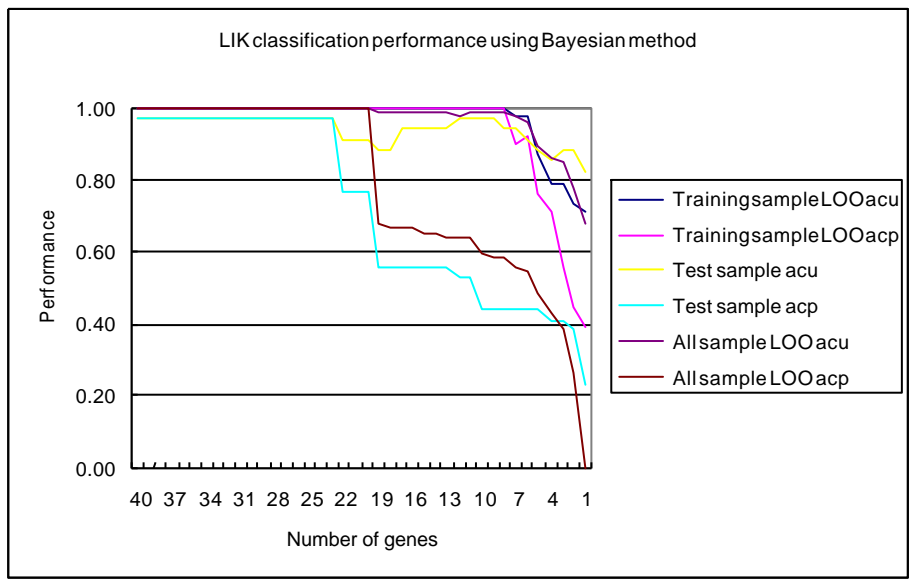

**Figure 4.** Performance of SVM and Naïve Bayesian classifiers built using genes selected according to LIK scores.

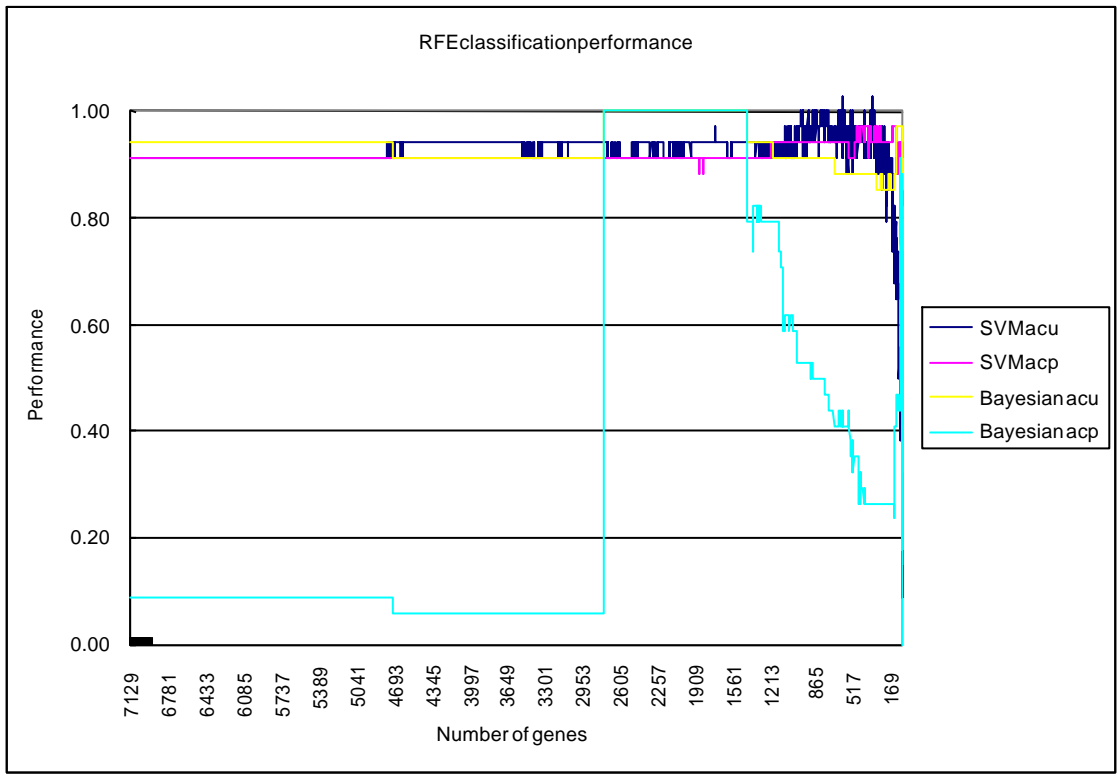

**Figure 5.** Classification performance of purely using RFE. Classification performance of SVM and Naïve Bayesian classifiers using genes selected by RFE starting from 7129 genes down to only one gene. The experimental setting was training test split and the performance measures were shown on the 34 test samples.

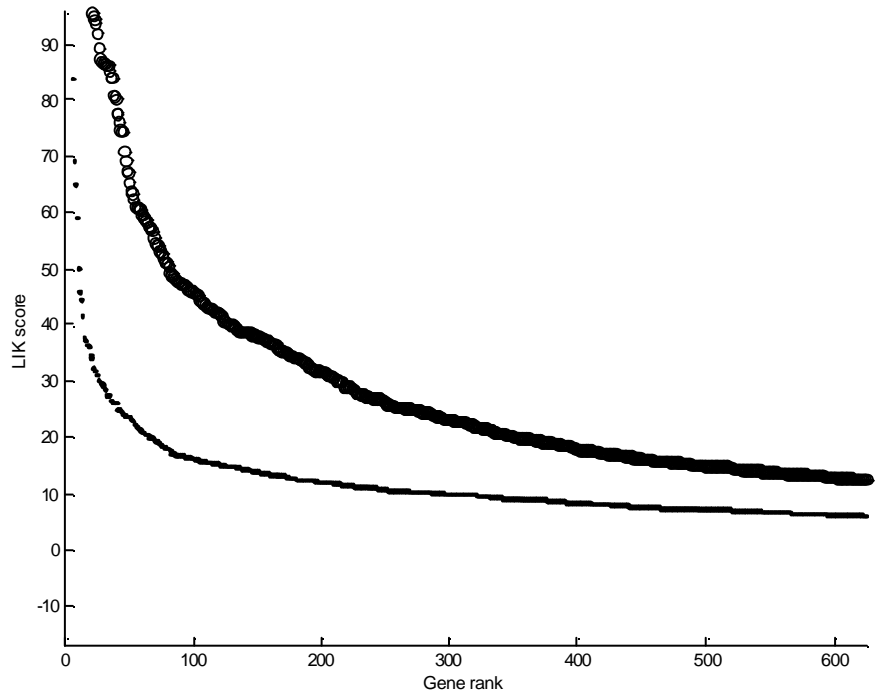

**Figure 6.** Sorted LIK scores of genes in the SRBCT dataset. Dots indicate $LIK_{EWS \to Non-EWS}$ scores and circles indicate $LIK_{Non-EWS \to EWS}$ scores.

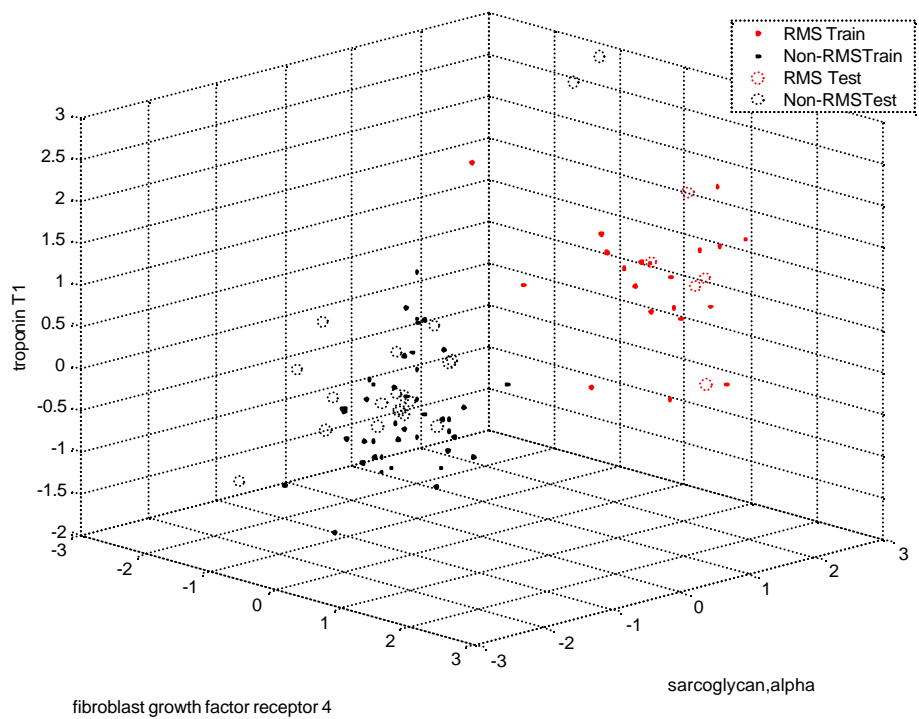

**Figure 7.** Plot of RMS and non-RMS samples. Plot of all 88 RMS and non-RMS samples according to the expression values of three of the four selected genes.

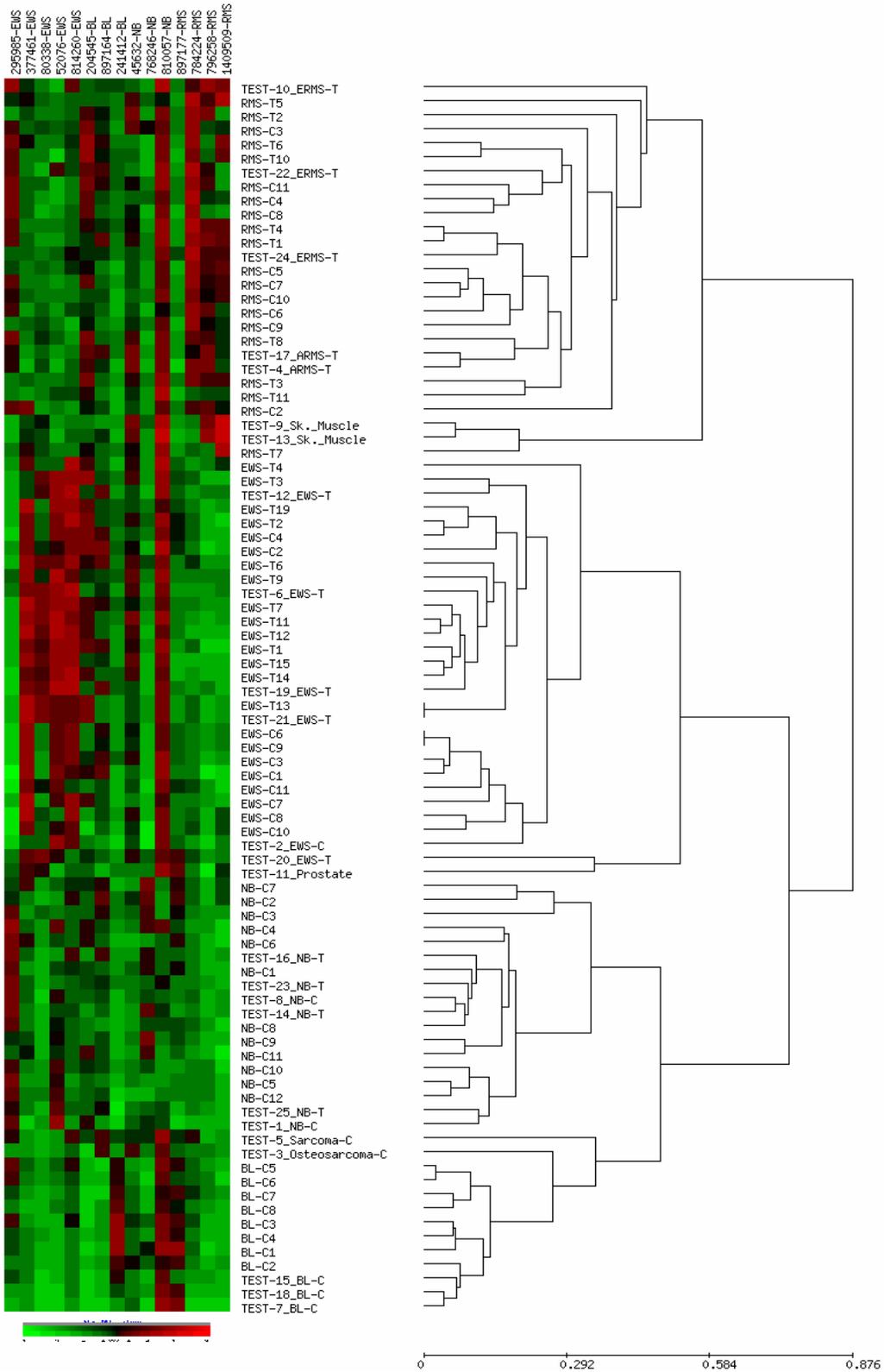

**Figure 8.** Hierarchical clustering of SRBCT samples with selected 15 genes.